\documentclass[11pt]{nature-withfigs}
\usepackage{amssymb,graphicx}

\newcommand\unit[1]{\ \mathrm{#1}}

\def\My{\unit{My}}

\def\Gy{\unit{Gy}}

\def\AU{\unit{AU}}
\def\km{\unit{km}}

\def\aap{Astron. Astrophys.}

\def\aj{Astron. J.}
\def\nat{Nature}

\def\apj{Astrophys. J.}
\def\apss{Astrophys. and Space Sci.}
\def\jrasc{J. R. Astron. Soc. Can.}

\title{A record of planet migration in the Main Asteroid Belt}

\author{David A. Minton$^{1}$ \& Renu Malhotra$^1$}

\begin{document}
\baselineskip=12pt

\maketitle

\begin{affiliations}
 \item Lunar and Planetary Laboratory, University of Arizona, 1629 E. University Blvd., Tucson, AZ 85716, USA\\ {-} \\
{\it Preprint: 23 December 2008}
\end{affiliations}

\begin{abstract}
The main asteroid belt lies between the orbits of Mars and Jupiter, but the region is not uniformly filled with asteroids.  
There are gaps, known as the Kirkwood gaps, in the asteroid distribution in distinct locations that are associated with orbital resonances with the giant planets\cite{Kirkwood:1867MA}; asteroids placed in these locations will follow chaotic orbits and be removed\cite{Wisdom:1987RSPSA}.  
Here we show that the observed distribution of main belt asteroids does not fill uniformly even those regions that are dynamically stable over the age of the solar system.  
We find a pattern of excess depletion of asteroids, particularly just outward of the Kirkwood Gaps associated with the 5:2, the 7:3, and the 2:1 jovian resonances. 
These features are not accounted for by planetary perturbations in the current structure of the solar system, but are consistent with dynamical ejection of asteroids by the sweeping of gravitational resonances during the migration of Jupiter and Saturn $\sim$4 gigayears ago.  
\end{abstract}

The Kirkwood gaps have been explained by the perturbing effects of the giant planets that cause dynamical chaos and orbital instabilities on very long timescales in narrow zones in the main asteroid belt\cite{Wisdom:1987RSPSA}, but thus far it has not been established how much of the main belt asteroid distribution is accounted for by planetary perturbations alone.
We compared the distribution of observed asteroids against a model asteroid belt uniformly populated in the dynamically stable zones.  
Our model asteroid belt was constructed as follows.  
Test particle asteroids were given eccentricity and inclination distribution similar to the observed main belt, but a uniform distribution in semimajor axis.  
We then performed a numerical integration for $4\Gy$ of the test particles' orbital evolution under the gravitational perturbations of the planets using a parallelized implementation of a symplectic mapping\cite{Wisdom:1991AJ,Saha:1992JRASC}.  
Details of the simulation can be found in the Supplementary Information. 

\begin{figure}
\includegraphics[width=11cm]{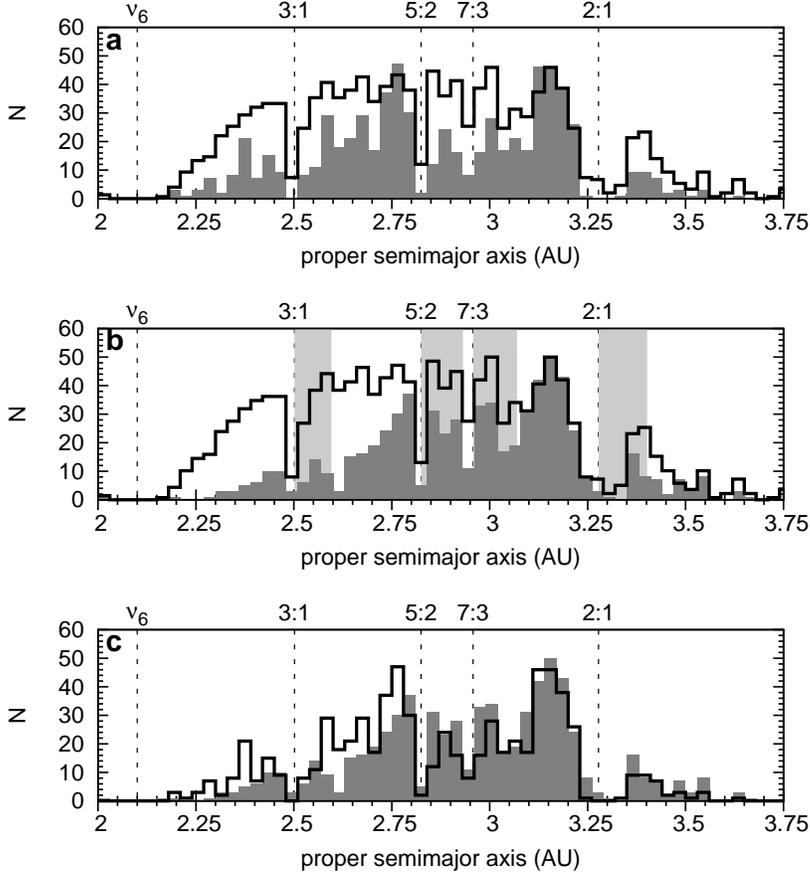}
\caption{{\bf Comparison of the observed main belt asteroid distribution with our simulated asteroid belt and results of the migration simulation.} 
{\bf a}, The solid line histogram is the distribution of asteroids remaining in our model asteroid belt at the end of the $4\Gy$ simulation in which the asteroid belt region was initially uniformly populated with test particles and the planets were in their current orbits.  
The shaded histogram is our observational comparison sample, which consists of the 690 asteroids with absolute magnitude $H<9.7$ (equivalent to diameters $D\gtrsim50$~km, assuming a visual geometric albedo of $0.09$), in the AstDys online data service\cite{Knezevic:2003AA}; see Supplementary Information for more details. 
The model asteroid belt (solid line) was normalized by multiplying all bins by a constant such that the value of the most-populous model bin equaled that of its corresponding bin in the observations. 
The current positions of the $\nu_6$ secular resonance and the strong jovian mean motion resonances associated with the major Kirkwood gaps are shown.  
{\bf b}, The solid line is the initial distribution of test particles in the simulation with migrating planets.  
The shaded histogram is the distribution of test particles remaining at the end of the 100 My migration simulation. 
The post-migration test particle bins were normalized by multiplying all bins by a constant value such that the value of the most-populous bin equaled that of its corresponding bin in the initial conditions.
The planet migration history followed the form of Eq.~\ref{e:migrationa}.  
The grey-shading indicates regions swept by the strong jovian mean motion resonances.
{\bf c}, Comparison of the model asteroid belt subjected to planet migration and the observed asteroid belt.  
The solid line is the distribution of observed large asteroids (the same as the shaded histogram of Fig.~1a).   
The shaded histogram is the distribution of test particles remaining at the end of the 100 My migration simulation (the same as the shaded histogram of Fig.~1b).  

\label{f:depleted-adist}}
\end{figure}

We sorted the surviving particles into semimajor axis bins of width $0.03\AU$.
We compared the model asteroid belt with the observed asteroid belt, as shown in Fig.~\ref{f:depleted-adist}a.  
We find that the observed asteroid belt is overall more depleted than the model can account for, and there is a particular pattern in the excess depletion:
there is enhanced depletion just exterior to the major Kirkwood gaps associated with the 5:2, 7:3, and 2:1 mean motion resonances (MMRs) with Jupiter (the regions spanning $2.81$--$3.11\AU$ and $3.34$--$3.47\AU$ in Fig.~\ref{f:depleted-adist}a);  
the regions just interior to the 5:2 and the 2:1 resonances do not show significant depletion (the regions spanning $2.72$--$2.81\AU$ and $3.11$--$3.23\AU$ in Fig.~\ref{f:depleted-adist}a), but the inner belt region (spanning 2.21--2.72 AU) shows excess depletion.

The above conclusions about the patterns of depletion are based on our model asteroid belt which assumes uniform initial population of the dynamically stable zones. 
It is conceivable that the discrepancies between the model and the observations could be due to a non-uniform initial distribution of asteroids.
However, the particular features we find cannot be explained by appealing to the primordial distribution of planetesimals in the solar nebula, nor to the effects of the mass depletion that occurred during the planet formation era (see Supplementary Information).  
As we show below, they can instead be readily accounted for by the effects of giant planet migration in the early history of the solar system.

There is evidence in the outer solar system that the giant planets -- Jupiter, Saturn, Uranus and Neptune -- did not form where we find them today.  
The orbit of Pluto and other Kuiper Belt Objects (KBOs) that are trapped in mean motion resonances with Neptune can be explained by the outward migration of Neptune due to interactions with a more massive primordial planetesimal disk in the outer regions of the solar system\cite{Malhotra:1993Nature,Malhotra:1995AJ}.  
The exchange of angular momentum between planetesimals and the four giant planets caused the orbital migration of the giant planets until the outer planetesimal disk was depleted of most of its mass, leaving the giant planets in their present orbits\cite{Fernandez:1984Icarus,Hahn:1999AJ,Tsiganis:2005Nature}.  
As Jupiter and Saturn migrated, the locations of mean motion and secular resonances swept across the asteroid belt, exciting asteroids into terrestrial planet-crossing orbits, thereby greatly depleting the asteroid belt population and perhaps also causing a late heavy bombardment in the inner solar system\cite{Liou:1997Science,Levison:2001Icarus,Gomes:2005Nature,Strom:2005Science}.  

We performed a computer simulation to test the hypothesis that the patterns of asteroid depletion inferred from Fig.~\ref{f:depleted-adist}a are consistent with planet migration. 
We used a total of 1819 surviving test particles from the previous $4\Gy$ simulation as initial conditions for a simulation with migrating planets. 
For the purposes of this simulation, we applied an external tangential force on each of the planets to simulate their orbital migration, so that a planet's semimajor evolved as follows\cite{Malhotra:1993Nature}:
\begin{equation}
	a(t)=a_0+\Delta a\left[1-\exp\left(-t/\tau\right)\right],
	\label{e:migrationa}
\end{equation} 
where $a_0$ is the initial semimajor axis, $\Delta a$ is the migration distance, and $\tau$ is a migration rate e-folding time.  
Jupiter, Saturn, Uranus, and Neptune had initial semimajor axes displaced from their current values by $\Delta a=+0.2$, $-0.8$, $-3.0$, and $-7.0\AU$, respectively;
these values are consistent with other estimates of Jupiter's and Neptune's migration distances\cite{Fernandez:1984Icarus,Malhotra:1995AJ,Franklin:2004AJ,Tsiganis:2005Nature}, but Uranus' and Saturn's migration distances are less certain. 
We used  $\tau=0.5\My$, which is near the lower limit inferred from Kuiper belt studies\cite{Murray-Clay:2005ApJ}.  
After $100\My$ of evolution under the influence of migrating planets, the 687 surviving test particles in the simulation were sorted and binned.  The distribution of the survivors is shown in Fig.~\ref{f:depleted-adist}b. 

We see directly that, in contrast with Fig.~\ref{f:depleted-adist}a, the asteroid belt distribution produced by the planet migration model matches qualitatively quite well the distribution of the observed asteroids (Fig.~\ref{f:depleted-adist}c).  
The patterns of excess depletion that we noted in Fig.~\ref{f:depleted-adist}a are explained well by the effects of the orbital migration of Jupiter and Saturn:  the regions within the sweep zones of the 5:2, 7:3, and 2:1 Jovian MMRs show enhanced depletion; the migration model also accounts for the excess depletion in the inner belt. 

\begin{figure}
\includegraphics[width=14cm]{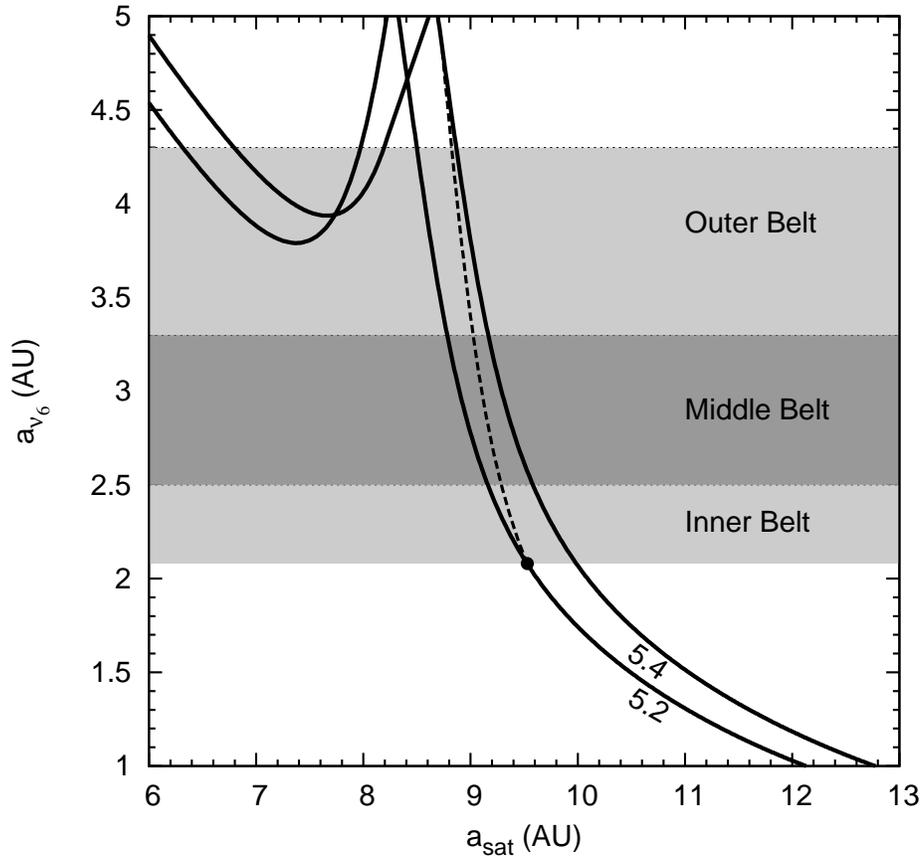}
\caption{{\bf The location of the $\nu_6$ secular resonance as a function of Saturn's semimajor axis.}  
This is  calculated using linear secular theory, with a correction for the 
effects of the (n+1):n MMRs, up to and including n=8, 
between Jupiter and Saturn\cite{Malhotra:1989AA}.  
The current location of the $\nu_6$ is designated with a dot.  
The labeled solid lines represent the calculation based on different fixed values of Jupiter's semimajor axis (given in AU).  
The dashed line is the path of the $\nu_6$ resonance in our migration simulation in which Jupiter and Saturn migrated $-0.2$ and $+0.8\AU$, respectively.\label{f:anu6vsasat}}
\end{figure}

Of note is that the inner asteroid belt region ($2.15$--$2.81\AU$) is somewhat more depleted in the migration simulation than in the observations.  
The majority of depletion from this region found in our migration simulation is due to the sweeping $\nu_6$ secular resonance. 
This powerful resonance removes asteroids from the main belt by secularly increasing their eccentricities to planet-crossing values\cite{Murray:1999SSD}.  
The maximum eccentricity of an asteroid disturbed by the passage of the $\nu_6$, and thereby the degree of asteroid depletion, is related to the sweeping speed: the slower the sweeping the more the depletion\cite{Heppenheimer:1980Icarus}.  
The distances the planets migrate determine the ranges in asteroids' semimajor axes that are affected by the sweeping.  
In our simulation, as Jupiter and Saturn migrated, the $\nu_6$ secular resonance swept inward across the entire main asteroid belt to its present location at $\sim 2.1\AU$, as shown in Fig.~\ref{f:anu6vsasat}.  
But because the $\nu_6$ resonance location is such a steep function of Saturn's semimajor axis (see Fig.~\ref{f:anu6vsasat}), for even modest proposed values of Saturn's migration distance, all of the asteroid belt is affected by the passage of this resonance.   
Thus, the overall level of depletion of the asteroid belt is most strongly dependant on the speed of planet migration, and only secondarily on the migration distance.  
Because we used an exponentially decaying migration rate for the giant planets, the $\nu_6$ resonance sweeping rate decreased as it approached its current location, thereby causing relatively greater asteroid depletion in the inner belt.  
Thus, the small but noticeable differences between the model and the observations in Fig.~\ref{f:depleted-adist}c are sensitive to the details of the time history of the planet migration speed. 

We note that our model asteroid belt lost $62\%$ of its initial pre-migration population, but the actual asteroid belt may have lost as much as $\sim90$--$95\%$ of asteroids due to migration\cite{OBrien:2007Icarus}.  
Because the overall level of asteroid depletion is particularly sensitive to the speed of planet migration, detailed exploration of parameters of the planet migration model and comparison with observations of main belt asteroids may provide strong quantitative constraints on planet migration.

\medskip
\noindent References



\begin{addendum}
 \item The authors acknowledge research funding from NASA and NSF.
 \item[Competing Interests] The authors declare that they have no
competing financial interests.
 \item[Correspondence] Correspondence and requests for materials should be addressed to\\ D.A.M. (email: daminton@lpl.arizona.edu).
\end{addendum}

\newpage
\newpage

\clearpage

\section*{Supplementary Information}
We restricted our observational comparison sample to asteroids with absolute magnitude $H<9.7$ (equivalent to diameters $D\gtrsim50$~km, assuming a visual geometric albedo of $0.09$) for three reasons. 
First, the Yarkovsky effect is not effective for large asteroids\cite{Farinella:1998Icarus}, so the only forces that have perturbed large asteroids over the solar system's history are gravitational and collisional.  
Second, the main asteroid belt is observationally complete for asteroid absolute magnitudes of at least $H\lesssim13$ ($D\gtrsim 11\km$ assuming a visual albedo of $0.09$)\cite{Jedicke:2002ASTIII}.  
Third, it has been estimated that the collisional destruction lifetimes of $D\gtrsim50\km$ asteroids are on the order of the age of the solar system\cite{Geissler:1996Icarus,Cheng:2004Icarus}.  
Therefore the population of asteroids with $H<9.7$ has not undergone appreciable collisional evolution since the end of the era of planet migration and our comparison sample is observationally complete. 

Our model asteroid belt was constructed as follows.  
A total of 5760 test particle asteroids were given initial conditions with both proper eccentricity and proper inclination distributions similar to the observed main belt, but a uniform distribution in proper semimajor axis.  
 The planets' masses were linearly increased from zero to their actual values over a period of $5\My$. This was done within an N-body simulation of the system of the Sun, the four outer giant planets and the test particle asteroids.
 Consequently the initial osculating elements of the test particles (when the planets masses were zero) became proper elements (as the planets became massive).  
 
We then performed a numerical integration for 4 Gyr of the test particles' orbital evolution under the gravitational perturbations of the four outer giant planets .  
The test particles were considered lost if they passed within 1 Hill radius of a planet (approximately the distance where the planet's influence on a test particle becomes comparable to the Sun's$^{16}$, because the subsequent orbital evolution is strongly chaotic), passed within the orbital radius of Mars, or reached a heliocentric distance greater than $100\AU$. 
All our numerical simulations were done with a parallelized implementation of a second-order mixed variable symplectic mapping (the Wisdom-Holman Method)$^{3,4}$.  In our simulations, the planets perturbed each other and perturbed the test particles, but the test particles had no effect on the planets. 
This is a reasonable approximation because the total mass in the asteroid belt is only $\sim10^{-6}$ of the total mass of the perturbing planets\cite{Krasinsky:2002Icarus}.
We also note that we did not include the terrestrial planets in our simulation, but their main effects were included by having an inner cutoff distance at the orbit of Mars for the removal of asteroids. We did perform a separate 260 Myr simulation that included explicitly the terrestrial planets; this is a much more computationally expensive simulation, hence its relatively short time span; we found very similar results, confirming that our results are robust.

The depletion patterns in Fig.~1a may be attributed to a number of plausible causes.  
Any mechanism invoked to explain the depletion must account for the features seen in Fig.~1a, namely that there is enhanced depletion in regions just outward of the major Kirkwood gaps.  
Here we explain why these features cannot be accounted for by appealing to the primordial distribution of planetesimals in the solar nebula, or to the effects of the first mass depletion event that occurred during the planet formation era.

First, we consider whether the depletion may be a reflection of the primordial mass distribution of planetesimals in the solar nebula, which is related to the surface mass density of the nebular disk, usually expressed as a decreasing function of heliocentric distance, $\Sigma\propto r^{-p}$.  
The ``minimum mass solar nebula'' (MMSN) model, which is derived by spreading out the masses of each planet in an annulus centered on its presumed initial semimajor axis and then fitting a surface mass density function, estimates that the index $p\approx 1.5$ if the giant planets are assumed to have formed in their initial locations, or $p\approx 2.2$ if the giant planets are assumed to have formed in a more compact configuration\cite{Weidenschilling:1977ApSS,Hayashi:1981PTPS,Desch:2007ApJ}.  
If mass is equally distributed among planetesimals of similar size in the asteroid belt region, then the number distribution of planetesimals is $N\propto r^{-p+1}$.  
Fig.~1a indicates that in the region between the inner edge of the asteroid belt and the 5:2 resonance, the number distribution of asteroids within the dynamically stable regions is an increasing function of heliocentric distance; neglecting the enhanced depletion adjacent to the Kirkwood gaps (the shaded regions in Fig.~1b), the overall depletion is approximately linear with $r$ in the region from $2.25\AU$ to $2.8\AU$, implying that $N\propto r$.  
Therefore a nebular surface mass density in accord with the MMSN model is inconsistent with that required to produce the observed asteroid distribution, and furthermore cannot account for the enhanced depletion adjacent to the Kirkwood gaps.

Second, we consider whether the asteroid distribution may have been produced during the planet formation era, due to either secular resonance sweeping related to the depletion of the solar nebula$^{17}$ or by gravitational perturbations from embedded planetary embryos in the asteroid belt region\cite{Wetherill:1992Icarus}.  
The sweeping secular resonance model may account for depletion in the inner asteroid belt seen in Fig.~1a if the rate of sweeping was an increasing function of semimajor axis.  
In the embedded planetary embryo model, the depletion in the inner asteroid belt could be accounted for if the scattering process was more efficient near the inner edge with decreasing efficiency outward.  
Thus the overall depletion trend does not contradict either model. 
However, neither model can readily account for the observed enhanced depletion adjacent to the Kirkwood gaps.

\section*{Supplementary References}

\end{document}